\documentclass{jpsj3}

\title{Optical detection of magnetic orders in HgCr$_2$O$_4$ frustrated spin magnet under pulsed high magnetic fields}

\author{Daisuke NAKAMURA, Atsuhiko MIYATA\thanks{Present address: Laboratoire National des Champs Magnetiques Intenses-Toulouse,  143, av. de Rangueil 31400 Toulouse, France}, Hiroaki UEDA\thanks{Present address: Department of Chemistry, Graduate School of Science, Kyoto University, Kyoto 606-8502, Japan.}, Shojiro TAKEYAMA\thanks{E-mail: takeyama@issp.u-tokyo.ac.jp}}
\inst{Institute for Solid State Physics, University of Tokyo, 5-1-5, Kashiwanoha,
Kashiwa, Chiba, 277-8581, Japan}
\abst{A magneto-optical survey was conducted for HgCr$_2$O$_4$ powder samples under pulsed high magnetic fields of up to 55 T. Intensity changes in magnetic fields observed for the exciton-magnon-phonon optical transition spectra coincide well with those of magnetization, lattice distortion from X-ray diffraction, and electron-magnetic resonances. The last-ordered phase was detected prior to the fully polarized magnetic phase, similarly to the other chromium spinel oxide, ZnCr$_2$O$_4$ and CdCr$_2$O$_4$.}

\kword{Magnetization,  pulsed high magnetic field, geometrically frustrated magnet, magneto-optical spectra }

\begin{document}
\maketitle

\section{Introduction}
HgCr$_2$O$_4$ is one of the chromium spinel oxides $A$Cr$_2$O$_4$ ($A$ = Mg, Zn, Cd, Hg), comprising a three-dimensional pyrochlore antiferromagnet with corner-sharing tetrahedra, and shows typical characteristics of geometrical spin frustration, represented by the large difference between the N\'{e}el and Weiss temperatures. It is well known that strong underlying spin-lattice coupling plays an important role in the magnetic properties of this magnetic-frustration system. When subjected to strong external magnetic fields, successive magnetic phase transitions take place accompanied by discontinuous lattice distortions \cite{tanaka}. The remarkable half-magnetization plateau phase \cite{hueda} and other antiferromagnetic phases realized in magnetic fields have been understood using a bilinear-biquadratic model that takes into account the spin-lattice interactions, as described in the theory developed by Penc et al. \cite{kpenc}

On decreasing atomic size going from Hg to Zn or Mg at the $A$ cation site in the pyrochlore lattice, much stronger magnetic fields are required to reveal all of the rich magnetic phases up to fully saturated magnetization, owing to the increased nearest-neighbor antiferromagnetic exchange interaction. 
Strong magnetic fields above 100 T, which could be used for solid-state physics measurements, have only been generated by magnet-coil destructive methods in an extreme environment, such as the single-turn coil and the electro-magnetic flux compression techniques, and are associated with relatively large instruments with their operation limited to microsecond time duration. 
Our group has explored an optical detection technique to study magnetic phases in extremely high magnetic fields and the methods have been applied to CdCr$_2$O$_4$, \cite{ekojima,amiyata} ZnCr$_2$O$_4$, \cite{amiyata2,amiyataprl,amiyatajpsj12} and MgCr$_2$O$_4$. \cite{amiyatajpsj14} 
The optical method was found to be a new and useful tool for detecting a magnetic phase that is hidden in conventional magnetization data and has revealed a new magnetic phase (described as the last ordered phase, LOP) prior to the fully polarized moment state. \cite{amiyata} 
The novel phase was concluded to be a phase corresponding to a magnetic `superfluid state', inferred from an analogy to symmetry breaking of the quantum phases in $^4$He as proposed by Matsuda and Tsuneto \cite{matsuda} and Liu and Fisher. \cite{Liu} 
The magnetic superfluid state is a state in which magnetic ordering takes place in the plane perpendicular to the magnetic field. Details of the spin structure of the novel phase could only be determined with comprehensive study based on other probes and measurements. However, the magnetic-field strengths ($B$) at which the LOP is realized are extremely large and methods of measurement probes are quite restricted.

We address two questions. One is whether the LOP observed in other chromium oxides also exists in HgCr$_2$O$_4$. The second is how the optically detected signals relate to the other measurements, such as X-ray diffraction and electron magnetic resonance (EMR). This paper focuses on magneto-absorption spectra appeared as an exciton-magnon-phonon (EMP) transition in HgCr$_2$O$_4$ in magnetic fields of up to 55 T. The results are discussed based on a comparison with those obtained from magnetization, \cite{huedaprb06} X-ray diffraction, \cite{tanaka} and EMR. \cite{kimura}

\section{Experiments}
%
\begin{figure}
\includegraphics[width=10 cm]{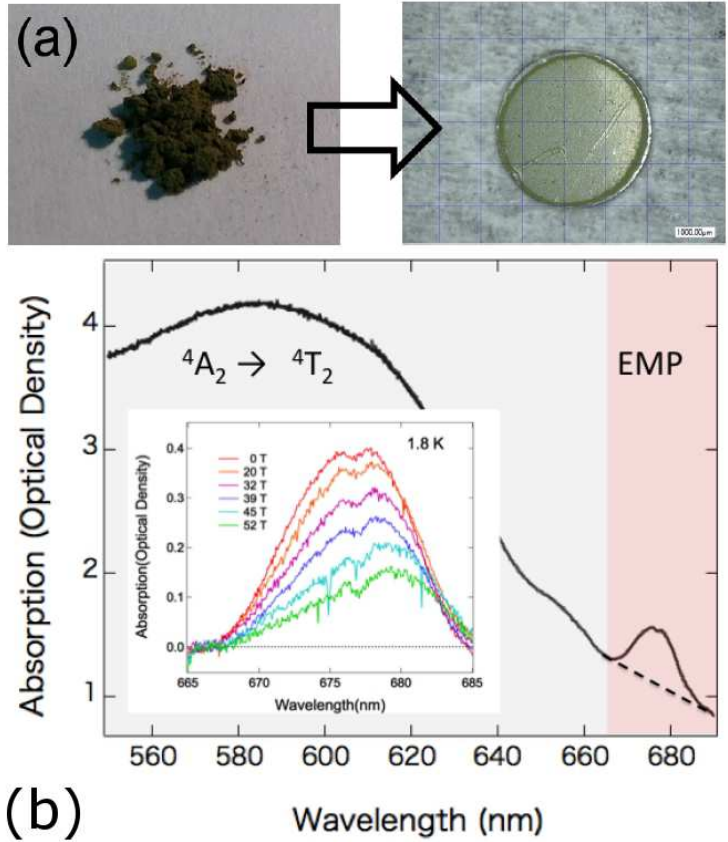}


\caption{\label{fig:Fig1ab}
(Color online)(a) Polycrystalline powder HgCr$_2$O$_4$ and a disk with the powder embedded in a Stycast resin synthesized in a manner described in the text. 
(b) Absorption spectra obtained
at 1.8 K in a region of wavelengths where $^{4}A_2-^{4}T_2$ and EMP transitions are observed in polycrystalline powder  HgCr$_2$O$_4$ embedded in Stycast resin.
}
\end{figure}
Polycrystalline powder of HgCr$_2$O$_4$, which was synthesized using thermal decomposition, \cite{huedaprb06} was dispersed in Stycast 1266B (liquid) catalyst with a weight ratio of the powder to the resign of 1:4. First, the B catalyst with dispersed powder was blended into the Stycast 1266A resin after confirmation of complete and uniform mixture of the powder with an assist of an ultrasonic disperser. The blended components were then dropped on a quartz disk (4 mm diameter and 0.4 mm thickness) and were dried one day in the air. The solidified mixture was then polished to a disk form with its thickness of approximately 50 $ \mu $m as shown in Fig.1 (a). Pulsed magnetic fields were generated from a nondestructive-type pulsed magnet which is capable of generating a magnetic field of up to 55 T with a half-period of 35 ms. A Xe-flash lamp was used as a light source. Optical fibers were used to guide the light in and out from the sample situated at the center of the magnet, and also to the polychromator equipped with an ICCD (image-intensified-charge-coupled device). The ICCD optical gate was open for 5 ms at the peak of the magnetic field so that the transmission light was recorded on the CCD arrays at each value of the magnetic field. The error in the magnetic field during the optical exposure time was approximately 2 T at the peak of 55 T. The sample temperature was kept at 1.8 K in a liquid helium cryogenic container. 

\section{Results}

Figure 1(b) displays the absorption spectra of a disk sample prepared in the manner described above. From the analogy of the spectra measured for CdCr$_2$O$_4$ \cite{amiyata} and ZnCr$_2$O$_4$, \cite{amiyatajpsj12} a broad but intense absorption peak at 590 nm is assigned as arising from the $^{4}A_2-^{4}T_2$ intra-d-band transitions (also see ref. 14). The small peaks that appear around 675 nm are attributed to the EMP transitions. The intra-d-band absorption peak ($^{4}A_2-^{4}T_2$) is much broader than those observed in CdCr$_2$O$_4$ and ZnCr$_2$O$_4$, probably because of the polycrystalline powder characteristics having associated stress and surface disorders. 
The EMP absorption spectra show decreasing peak intensity with increasing magnetic field, as seen in Fig. 2 (a), where the integrated intensity of the peak is plotted against magnetic field. As noted from this plot, the rate of intensity decrease is not constant. There are several ranges of magnetic-field strength for which the rate of decrease changes. The intensity is almost constant up to 13 T, and then decreases until 27 T, where the rate of decrease slows down. There are kinks at 35 T, 45 T, and 48 T indicative of the changes of the decreasing rate. The magnetic-field strengths at which the rate changes exhibit very good correspondence with magnetization steps. 
Figure 2(b) is a plot of a magnetization $M$ and its derivative d$M$/d$B$ in magnetic fields measured at a temperature of 1.8 K,\cite{huedaprb06} where the successive field-induced phase transitions have been reported up to a full saturation of the moment. Starting from the antiferromagnetic phase, $M$ increases linearly up to 10 T ($B_{c1}$), then exhibits a sudden jump associated with hysteresis before entering into the half-magnetization plateau phase (a collinear three-up, one-down spin configuration) above 13 T. This plateau continues up to 27 T ($B_{c2}$), above which $M$ increases linearly again up to 35 T ($B_{c3}$), where it exhibits a canted 3:1 phase. Finally, M shows a sudden increase with a small hysteresis jump prior to a gradual saturation of the magnetization moment of 3$\mu _{B}$/Cr$^{3+}$ in magnetic fields between 37 T and 45 T.

\section{Discussion}
As seen in Fig. 2(b), the transition points observed in $M$ are associated with a finite width in magnetic fields, which are well recognized using the peak width in $dM/dB$, and are illustrated by a hatching zone. Note that the width at the $B_{c1}$ transition is very wide in comparison to the other chromium spinels, as has already been pointed out in ref. 12. They speculated that this is due to the effect of disorder in the nonmagnetic cation site (the Hg site) in the polycrystalline samples. It is readily evident from Fig. 2(a) and (b) that the kinks observed in the intensity of the EMP transition spectra always occur in these hatching zones up to $B_{c3}$, i.e., there is always a good correspondence between the magnetic field at which the kink takes place and those in the transition field in $M$.

Details about successive distortions of the pyrochlore lattice with an application of magnetic fields in HgCr$_2$O$_4$ were revealed using synchrotron X-ray diffraction with a specially designed pulse magnet in magnetic-field strengths up to 38 T. \cite{tanaka} Figure 2(c) is a plot of a change in the lattice parameter, $\Delta a(B)=a(B)-a(B=0) $ in magnetic fields, which is reproduced from ref. 1. We note the following two major results. The lattice constant changes up to $B_{c1}$, whereas the EMP intensity stays constant. In the plateau magnetic phase, the lattice constant stays constant, while the EMP intensity shows a linear decrease. These contrasting behaviors of the EMP intensity are evidence that the EMP intensity predominantly reflects the state of the magnons (the magnon density of states) but is less influenced by the changes in the lattice, in spite of the tight involvement of both phonons and magnons for the EMP optical transition.

According to the discussion of Tanaka et al.,\cite{tanaka} all the bond lengths elongate equally in the pyrochlore lattice in the fully polarized phase in magnetic fields above $B_{c3}$. Hence, the lattice constant should be constant above $B_{c3}$, and $M$ is regarded as showing full saturation accompanied by gradual increases due to an effect of finite temperature. However, the EMP intensity shows a kink at $B_{c3}$ and continues to decrease followed by another kink at 45 T ($B_{c4}$). The anomaly observed in the EMP transition intensity between $B_{c3}$ and $B_{c4}$ suggests the existence of a novel phase in the region that is believed to be a fully polarized phase. A similar phase has already been reported by our group, inferred from anomalies in the optical absorption spectra of ZnCr$_2$O$_4$ \cite{amiyataprl,amiyatajpsj12} and CdCr$_2$O$_4$. \cite{amiyata}

Let us note here the discussion by Kimura et al. who have conducted EMR together with magnetization measurements in strong pulsed magnetic fields. \cite{kimura} Firstly, they observed in $dM/dB$ an evident but weak peak around 39 T beside the main sharp peak at 35 T ($B_{c3}$), which was attributed to a possible new phase between the canted (3:1) ferromagnetic phase and the fully polarized phase. A similar but rather broad peak around 82 T in $dM/dB$ was also observed by our group for CdCr$_2$O$_4$ measured using a single-turn coil.\cite{amiyata} We ascribed this anomaly to a transition to a novel phase, i.e., the LOP, in which a magnetic `superfluid' state in a boson picture (such as either an umbrella-like spin structure or a spin-nematic state) is realized. Second, rapid disappearance of the EMR $w_u$ mode at 36 T, corresponding to a sharp transition peak in $dM/dB$, is indicative of the last field-induced lattice transition to a high-symmetry cubic structure. This scenario is also consistent with those in the discussion above by Tanaka et al. Therefore, the LOP transition from the 3:1 canted phase is first-order and accompanied by the lattice distortion transformed into a higher symmetry.

According to a finite-temperature Monte-Carlo simulation of carried out by Motome et al. for a magnetization in a classical pyrochlore lattice with incorporation of the bilinear and the biquadratic interactions,\cite{Motome} there appears in susceptibility a broad shoulder at the high magnetic field side of the sharp peak of the transition into the final fully polarized phase. This shoulder-like structure arises from the temperature-dependent broadening of the last first-order transition and never exhibits a peak structure as is observed in real experiments. Therefore, the small peak in $dM/dB$ magnetization data around 39 T that appears for HgCr$_2$O$_4$ is similarly associated with a phase transition to the LOP prior to the full-saturation phase, as has been observed in CdCr$_2$O$_4$ and ZnCr$_2$O$_4$.  A sudden decrease of the EMP intensity is observed above $B_{c4}$, where magnetization showed almost complete saturation. This sudden decrease suggests the existence of a finite gap for magnon excitation from the fully polarized state, which is regarded as the ground state in a magnon picture.

\section{Summary}

Optical absorption arising from an EMP transition observed in powdered HgCr$_2$O$_4$ was measured in magnetic fields up to 55 T. The EMP absorption, reflecting a magnon density of states, showed intensity changes corresponding to phase transitions consistent with magnetization, lattice distortions, and EMR successive anomalies reported previously. This study has revealed that a fourth phase, termed LOP, also exists in HgCr$_2$O$_4$ and is related to the magnetic superfluid state that has also been observed in CdCr$_2$O$_4$ and ZnCr$_2$O$_4$ between the canted (3:1) phase and the full-saturation phase.

\begin{figure}
\includegraphics[width=11 cm]{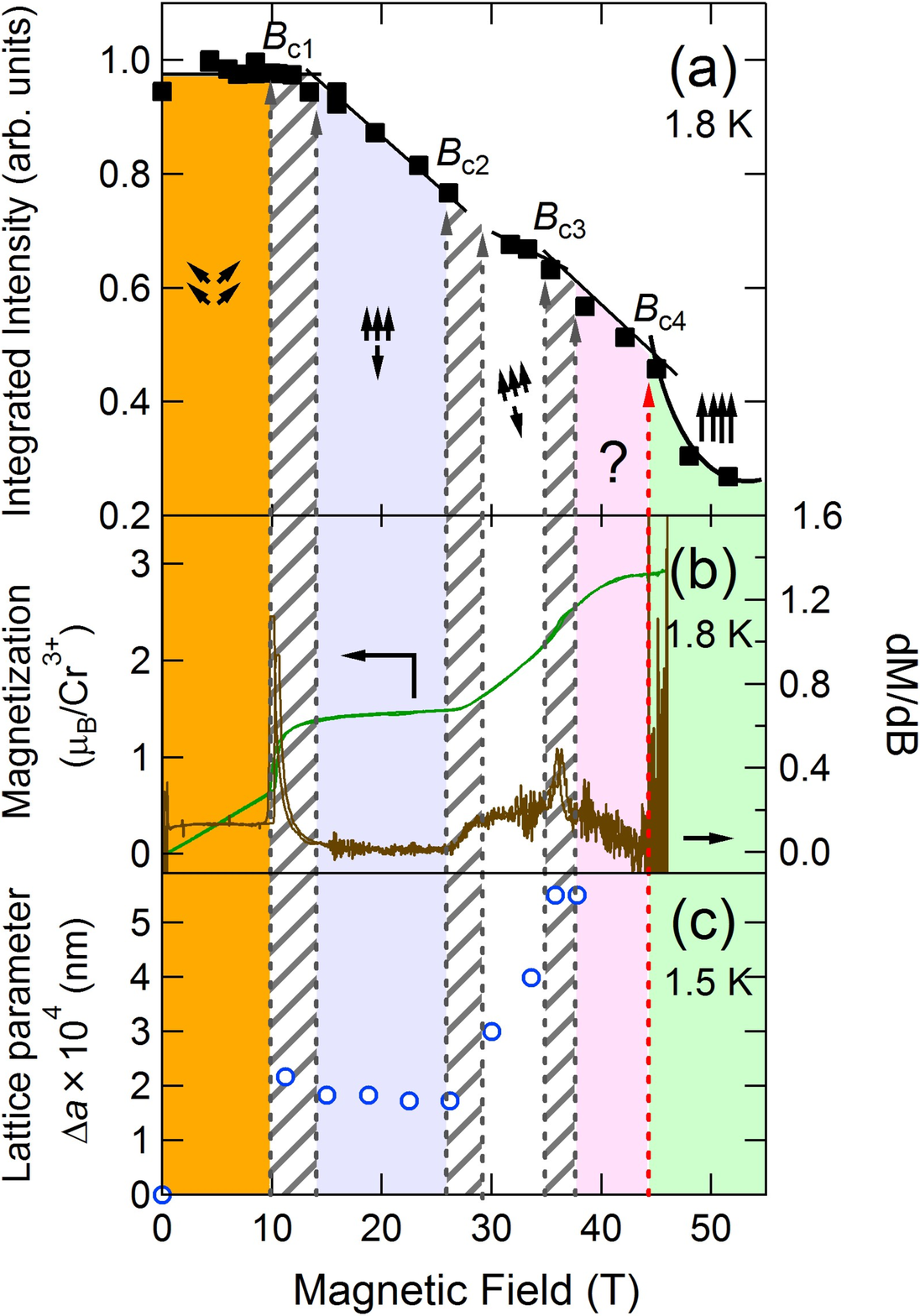}\\
\caption{
(Color online) (a) Changes in an integrated intensity of the EMP absorption peak in magnetic fields. (b) Magnetization $M$ and its derivative $dM/dB$ as a function of magnetic field, measured at 1.8 K, (ref.12) and (c) Change of the lattice parameter $\Delta a$ plotted against a magnetic field. The data (circle) are reproduced from Fig.2 in ref. 1.
}
\label{fig:Fig2abc}
\end{figure}

\section*{Acknowledgments}
The authors are obliged to Prof. K. Kindo for providing us with a non-destructive magnet.
A.M. thanks to a support of the Grant-in-Aid for JSPS Fellows.


\end{document}